\documentclass{aa}  
\usepackage{pictex}
\usepackage{graphicx}
\usepackage{color}
\usepackage{natbib}
\usepackage{subfigure} 
\bibpunct{(}{)}{;}{a}{}{,}
\usepackage{txfonts}

\newcommand{\om}{\Omega_\mr m}
\newcommand{\sig}{\sigma_8}
\newcommand{\C}{\mathbf C}

\newcommand{\vxi}{\vec \xi}

\newcommand{\vpi}{\vec \pi}

\newcommand{\map}{\langle M_\mr{ap}^2\rangle}

\newcommand{\rrE}{\ensav{\mathcal R \mathcal R_\mr E}}
\newcommand{\rrB}{\ensav{\mathcal R \mathcal R_\mr B}}
\newcommand{\xip}{\xi_+}
\newcommand{\xim}{\xi_-}

\newcommand{\mr}{\mathrm}

\newcommand{\tn}{\textnormal}
\newcommand{\be}{\begin{equation}}
\newcommand{\ee}{\end{equation}}

\newcommand{\nn}{\nonumber}
\newcommand{\beq}{\begin{eqnarray}}
\newcommand{\eeq}{\end{eqnarray}}
\newcommand{\ensav}[1]{\left\langle #1 \right\rangle}

\begin{document}

   \title{Measuring cosmic shear with the ring statistics}
\subtitle{}

   \author{T. Eifler \inst{1}, P. Schneider \inst{1} and E. Krause \inst{2,1}}
   \offprints{tim.eifler@astro.uni-bonn.de}

   \institute{1) Argelander-Institut f\"ur Astronomie, Universit\"at Bonn, Auf dem H\"ugel 71, D-53121 Bonn, Germany \\
   2) California Insitute of Technology, M/C 350-17, Pasadena, California 91125, USA}

   \date{}

\abstract
{Commonly used methods to decompose E- and B-modes in cosmic shear, namely the aperture mass dispersion and the E/B-mode shear correlation function, suffer from incomplete knowledge of the two-point correlation function (2PCF) on very small and/or very large scales. The ring statistics, the most recently developed cosmic shear measure, improves on this issue and is able to decompose E- and B-modes using a 2PCF measured on a finite interval.} 
{First, we improve on the ring statistics' filter function with respect to the signal-to-noise ratio. Second, we examine the ability of the ring statistics to constrain cosmology and compare the results to cosmological constraints obtained with the aperture mass dispersion. Third, we use the ring statistics to measure a cosmic shear signal from CFHTLS (Canada-France-Hawaii Telescope Legacy Survey) data.}
{We consider a scale-dependent filter function for the ring statistics which improves its signal-to-noise ratio. To examine the information content of the ring statistics we employ ray-tracing simulations and develop an expression of the ring statistics' covariance in terms of a 2PCF covariance. We perform a likelihood analysis with simulated data for the ring statistics in the $\om$-$\sig$ parameter space and compare the information content of ring statistics and aperture mass dispersion. Regarding our third aim, we use the 2PCF of the latest CFHTLS analysis to calculate the ring statistics and its error bars.}
{Although the scale-dependent filter function improves the S/N ratio of the ring statistics, the S/N ratio of the aperture mass dispersion is higher. In addition, we show that there exist filter functions which decompose E- and B-modes using a finite range of 2PCFs ($EB$-statistics) and have higher S/N ratio than the ring statistics. However, we find that data points of the latter are significantly less correlated than data points of the aperture mass dispersion and the $EB$-statistics. As a consequence the ring statistics is an ideal tool to identify remaining systematics accurately as a function of angular scale. We use the ring statistics to measure a E- and B-mode shear signal from CFHTLS data. } 
{}

\keywords{cosmology: theory - gravitational lensing - large-scale structure of the Universe - methods: statistical}

\maketitle
%

\section{Introduction}
Cosmic shear was first detected in 2000 \citep{bre00,kwl00,wme00,wtk00} and has progressed to a valuable source of cosmological information. Latest results \citep[e.g.,][]{wmh05,smw06,hmv06,ses06,het07,mrl07,fsh08} already indicate its great potential to constrain cosmological parameters, which will be enhanced  by large upcoming surveys like Pan-STARRS, KIDS, DES or Euclid.\\
An important step in a cosmic shear analysis is the decomposition into E- and B-modes, where, to leading order, gravitational lensing only creates E-modes. In principle, B-modes can arise from the limited validity of the Born approximation \citep{jsw00,hhw08} or redshift source clustering \citep{svm02}. Another possible source are astrophysical contaminations such as intrinsic alignment of source galaxies; \cite{ks03} show how to separate the cosmic shear signal from intrinsic alignment contaminations if redshift information is available. The strength of B-modes coming from these effects are examined through numerical simulations; although the results differ \citep[e.g.][]{hrh00,crit01,jing02}, the observed B-mode amplitude is higher than expected from the foregoing explanations. Shape-shear correlation \citep{hs04} is another astrophysical contamination which can cause B-modes. \cite{js08, js09} show how to exclude the contaminated scales, again using redshift information.\\
Most likely, B-modes indicate remaining systematics in the observations and data analysis, in particular they can result from an insufficient PSF-correction. The Shear TEsting Program (STEP) has significantly improved on this issue \citep[for latest results see][]{hwb06,mhb07}; still the accuracy of the ellipticity measurements must be improved further to meet the requirements of precision cosmology.\\
The identification of remaining systematics (B-modes) will be important especially for future surveys, where the statistical errors will be significantly smaller. Therefore, decomposing the shear field into E- and B-modes must not be affected from inherent deficits. The most commonly used methods for an E- and B-mode decomposition, the aperture mass dispersion and the E/B-mode shear correlation function, require the shear two-point correlation (2PCF from now on) to be known down to arbitrary small or up to arbitrary large angular separations, respectively. This is not possible in practice; as a consequence the corresponding methods do not separate E- and B-modes properly on all angular scales. A detailed analysis of this issue can be found in \cite{kse06} (hereafter KSE06).\\
Most cosmic shear analyses, e.g. \cite{mrl07} and \cite{fsh08} (hereafter FSH08), simulate 2PCFs from a theoretical model of $P_\kappa$ to account for the scales on which the 2PCF cannot be obtained from the data. This ansatz is problematic, since one explicitly assumes that the corresponding scales are free of B-modes. In addition, the assumed cosmology in the theoretical power spectrum can bias the results.\\
The ring statistics \citep[][hereafter SK07]{sk07} provides a new method to perform an E-/B-mode decomposition using a 2PCF measured over a finite angular range $[\vartheta_\mr{min};\vartheta_\mr{max}]$. In this paper we examine the ring statistics in detail; more precisely we improve the ring statistics' filter function with respect to its S/N ratio and examine its ability to constrain cosmological parameters. Furthermore, we construct a filter functions which has higher S/N ratio than the ring statistics but still decomposes E/B-modes with a 2PCF measured over a finite range. We will refer to this as $EB$-statistics.\\
Due to the fact that the ring statistics' data points show significantly lower correlation than data points of the aperture mass dispersion and the $EB$-statistics, it provides an ideal tool to identify remaining systematics in cosmic shear surveys depending on the angular scale. We employ the ring statistics to identify B-modes in the CFHTLS survey.\\
The paper is structured as follows: In Sect. \ref{sec:basics} we start with the basics of second-order cosmic shear measures, followed by the main concepts of the ring statistics in Sect. \ref{sec:ringstatistics}. We derive a formula to calculate the ring statistics' covariance from a 2PCF covariance in Sect. \ref{sec:covring} and also compare the correlation coefficients of ring statistics, aperture mass dispersion, and $EB$-statistics in this section. In the same section we examine the S/N ratio of the ring statistics and compare it to the other measures. More interesting than the S/N ratio however, is the ability of a measure to constrain cosmology. This, in addition to the S/N ratio, depends on the correlation of the individual data points. In order to quantify this accurately, we perform a likelihood analysis in Sect. \ref{sec:mapvsring} for the ring statistics, aperture mass dispersion and $EB$-statistics using data from ray-tracing simulations. The results of our analysis of CFHTLS data using the ring statistics are presented in Sect. \ref{sec:cfhtlsring} followed by our conclusions in Sect. \ref{sec:conc}. 
\section{Two-point statistics of cosmic shear}
\label{sec:basics}
In this section we briefly review the basics of second-order cosmic shear measures. For more details on this topic the reader is referred to \cite{bas01,svk02,svm02,wam03,mvv08}. \\
To measure the shear signal we define $\vec \vartheta$ as the connecting vector of two galaxy centers and specify tangential and cross-component of the shear $\gamma$ as
\be
\gamma_\mr t = - \mr{Re} \left( \gamma \mr e^{-2\mr i \varphi} \right) \qquad  \tn{and}  \qquad
\gamma_{\times} = - \mr{Im} \left( \gamma \mr e^{-2\mr i \varphi} \right) \;,
\ee 
where $\varphi$ is the polar angle of $\vec \vartheta$. The 2PCFs depend only on the absolute value of $\vec \vartheta$. They are defined in terms of the shear and can be related to the power spectra $P_{\mr E}$ and $P_{\mr B}$ \citep{svm02}
\beq
\label{eq:xifrome}
\xi_{\pm} (\vartheta) &\equiv& \langle \gamma_\mr t \gamma_\mr t  \rangle (\vartheta) \pm \langle \gamma_{\times} \gamma_{\times} \rangle (\vartheta) \\
\label{eq:xi+-}
&=&\int^{\infty}_0 \frac{\mr d\ell\;\ell}{2\pi} \, \mr J_{0/4}(\ell \vartheta)\, \,\left[P_\mr E(\ell) \pm P_\mr B(\ell)\right]\,,
\eeq
with $\mr J_n$ denoting the $n$-th order Bessel-function.\\ 
Starting from the 2PCF as the basic observable quantity, there exist several methods to decompose E-modes and B-modes, such as the E/B-mode shear correlation function or the aperture mass dispersion \cite[e.g.][]{cri02,svm02}. The latter can be calculated as 
\be
\label{eq:mapfromxi}
\langle M^2_{\mr{ap}/\bot} \rangle (\theta) = \frac{1}{2} \int_{0  }^{2\theta} \frac{\mr d \vartheta \, \vartheta}{\theta^2} \left[ \xi_+ (\vartheta) T_+ \left( \frac{\vartheta}{\theta} \right) \pm \xi_- (\vartheta) T_- \left( \frac{\vartheta}{\theta} \right) \right] \,.
\ee
The filter functions read
\beq
\label{eq:T_+}
T_+ (x) & = &  \left\{ \frac{6(2-15 x^2)}{5} \left[ 1 - \frac{2}{\pi} \arcsin{\left(\frac{x}{2}\right)} \right]+  \frac{x \sqrt{4-x^2}}{100 \pi} \left( 120  \right.\right. \nonumber \\
 &\,& \left. \left. + 2320x^2-754x^4+132x^6-9x^8 \right) \right\}  \mr{H}(2-x)\, ,  \\
\label{eq:T_-}
T_- (x) & = & \frac{192}{35 \pi} x^3 \left(1-\frac{x^2}{4} \right)^{7/2} \mr{H}(2-x) \,,
\eeq
with $H$ being the Heaviside step function. Decomposing E- and B-modes with the either the aperture mass dispersion or the E/B-mode shear correlation function requires that the 2PCF is either measured down to arbitrary small or large angular separation, respectively. For further details on this problem the reader is referred to KSE06. 
\section{The ring statistics}
\label{sec:ringstatistics}
\begin{figure}
 \includegraphics[width=9cm]{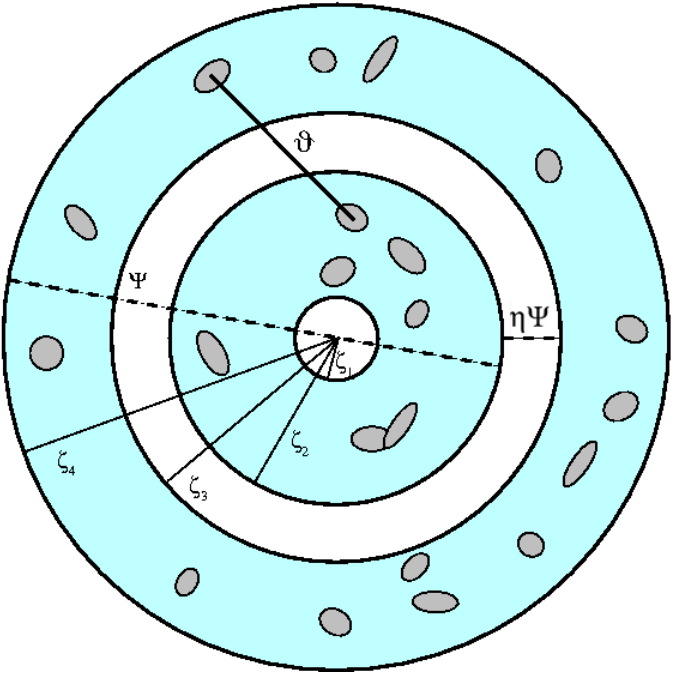}
   \caption{This figure illustrates the basic idea of the ring statistics and how it can be obtained from the 2PCF of cosmic shear. We measure the 2PCF of each galaxy in the inner ring with all galaxies in the outer ring. For a given argument of the ring statistics $\Psi$, the angular separation of the required 2PCFs extends over $\eta \Psi \leq \vartheta \leq \Psi$. The meaning of $\eta$ and its possible values are further explained in the text. The ring statistics is then calculated as an integral over the 2PCF with the filter functions $Z_\pm (\vartheta, \eta)$.}
         \label{fig:ringprinciple}
\end{figure}

\begin{figure*}
\includegraphics[width=17cm]{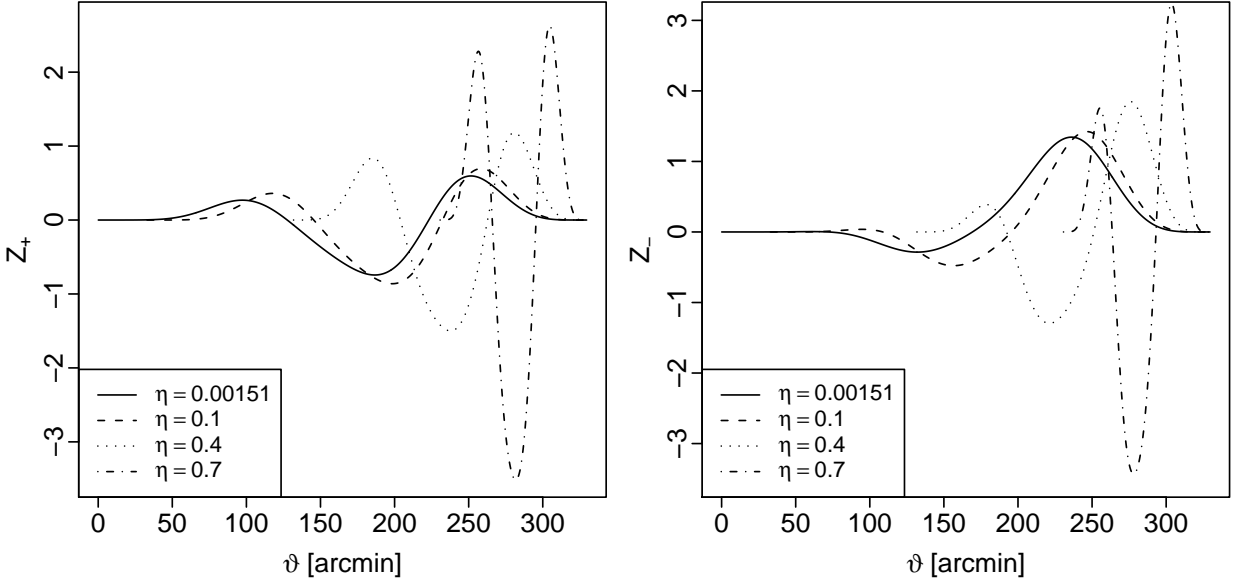}
\caption{This plot shows the filter functions $Z_+$ (\textit{left} ) and $Z_-$ (\textit{right}) depending on $\vartheta$ for four different choices of $\eta$: $\vartheta_\mr{min}/\Psi=0.00151$ (\textit{solid}), 0.1 (\textit{dashed}), 0.4 (\textit{dotted}), 0.7 (\textit{dotted dashed}).}
\label{fig:zfunc_CFHTLS}
\end{figure*}

To circumvent the aforementioned difficulties SK07 introduced the ring statistics whose second-order moments ($\rrE, \rrB$) decompose E- and B-modes properly using 2PCFs measured on a finite interval $[\vartheta_\mr{min};\vartheta_\mr{max}]$. The quantity $\rrE$ can be interpreted as the correlator of the shear measured from galaxy pairs which are located inside two concentric rings (see Fig. \ref{fig:ringprinciple}). Their annuli are chosen as follows: $\zeta_1 \leq \theta_1 \leq \zeta_2$ for the first ring and $\zeta_3 \leq \theta_2 \leq \zeta_4$ for the second. The rings are non-overlapping, i.e. $\zeta_i < \zeta_j$ if $i<j$. The argument of the rings statistics is named $\Psi=\zeta_2+\zeta_4$ and only 2PCFs with $\vartheta \leq \Psi$ enter in the calculation of $\rrE$. In addition, the ring statistics depends on a parameter $\eta$ quantifying the separation between outer and inner ring, i.e.  $\eta/ \Psi= \zeta_3 - \zeta_2$. In order to calculate the ring statistics properly from a set of 2PCFs within $[\vartheta_\mr{min};\vartheta_\mr{max}]$ it is required that $\Psi$ does not exceed $\vartheta_\mr{max}$ and that $\vartheta_\mr{min}/\Psi \leq \eta < 1$.\\ 
Following the derivation of SK07 the E- and B-mode decomposition of the ring statistics can be obtained from the 2PCF as 
\beq
\label{eq:ringe}
\rrE (\Psi)&=&\int_{\eta \Psi}^\Psi  \frac{\mr d \vartheta}{2\,\vartheta} \left[ \xip(\vartheta) \, Z_+(\vartheta,\eta) + \xim(\vartheta) \, Z_-(\vartheta,\eta)\right]\,, \\
\label{eq:ringb}
\rrB (\Psi)&=& \int_{\eta \Psi}^\Psi  \frac{\mr d \vartheta}{2\,\vartheta} \left[ \xip(\vartheta) \, Z_+(\vartheta,\eta) - \xim(\vartheta) \, Z_-(\vartheta,\eta) \right]\,.
 \eeq 
The functions $Z_\pm$ are defined in SK07; we plot them in Fig. \ref{fig:zfunc_CFHTLS} for four different $\eta$, i.e. $\vartheta_\mr{min}/\Psi=0.00151, 0.1, 0.4, 0.7$.\\
Similar to the case of the aperture mass dispersion, $\rrE$ can be related to the E-mode power spectrum. Inserting  Eq. (\ref{eq:xi+-}), into Eq. (\ref{eq:ringe}) gives
\be 
\label{eq:ringfrompowe}
\rrE (\Psi) = \int_0^\infty \frac{\mr d \ell \, \ell}{2 \pi} \, P_\mr E (\ell)\, \mathcal W_\mr E (\ell \Psi, \eta)
\ee
with 
\be 
\label{eq:Wfunction}
\mathcal W_\mr E(\ell \Psi, \eta)= \int_{\eta \Psi}^\Psi \frac{\mr d \vartheta}{2 \vartheta} \left[ J_0 (\ell \vartheta)\, Z_+(\vartheta, \eta)+ J_4 (\ell \vartheta)\, Z_-(\vartheta, \eta)\right] .
\ee  
When calculating $\rrE$ for different arguments $\Psi$, we distinguish two cases for $\eta$. It can be fixed to a specific value or it can vary according to $\Psi$, in particular $\eta=\vartheta_\mr{min}/\Psi$. We will refer to the latter case as a scale-dependent $\eta$. Here, the lower limit in the integrals of Eqs. (\ref{eq:ringe}) and (\ref{eq:ringb}) is equal to $\vartheta_\mr{min}$ which implies that all 2PCFs in the interval $[\vartheta_\mr{min}; \Psi]$ are included in the calculation. The choice of $\eta=\vartheta_\mr{min}/\Psi$ should give a higher S/N ratio compared to a fixed $\eta$ for the reason that more galaxy pairs are included which reduces the statistical noise. In SK07 the authors hold $\eta$ fixed; in order to obtain a high signal this implies that $\eta$ must be chosen as small as possible.\\
Choosing a fixed $\eta$ has a second disadvantage. The lower limit in the integrals Eqs. (\ref{eq:ringe}) and (\ref{eq:ringb}) cannot be smaller than $\vartheta_\mr{min}$, i.e. $\eta \Psi \geq \vartheta_\mr{min}$. Vice versa, this implies that $\Psi \geq \vartheta_\mr{min}/\eta$. Fixing $\eta$ to a small value (in order to increase the S/N ratio) implies that $\Psi$ is restricted to larger scales. This trade-off between S/N ratio and small-scale sensitivity can be overcome when relaxing the condition of a fixed $\eta$. 
\section{General E/B-mode decomposition on a finite interval}
The ring statistics described in the last section is the special case of a general E/B-mode decomposition. According to SK07 this general $EB$-statistics can be defined as
\beq
\label{eq:ebgeneral}
 E &=&  \frac{1}{2} \,\int_0^{\infty} \mr d \vartheta \, \vartheta \left[ \xi_+ (\vartheta) T_+ (\vartheta) + \; \xi_- (\vartheta) T_- (\vartheta) \right]  \,,\\
 B  &=&  \frac{1}{2} \, \int_0^{\infty} \mr d \vartheta \, \vartheta \left[ \xi_+ (\vartheta) T_+ (\vartheta) - \; \xi_- (\vartheta) T_- (\vartheta) \right]  \,. 
\eeq 
To provide a clean separation of E- and B-modes using a 2PCF measured over a finite interval, the following conditions must be fulfilled (see SK07 for the exact derivation). Starting from an arbitrary function $T_+(\vartheta)$, which is zero outside the interval $[\vartheta_\mr{min};\vartheta_\mr{max}]$, the constraints
\be
\label{eq:ebfinitecondition2}
\int_{\vartheta_{\min}}^{\vartheta_{\max}} \mr d \vartheta \, \vartheta T_+(\vartheta) = 0 =\int_{\vartheta_{\min}}^{\vartheta_{\max}} \mr d \vartheta \, \vartheta^3 T_-(\vartheta)
\ee
must hold. For a so constructed filter function  $T_+(\vartheta)$ a corresponding filter function $T_-(\vartheta)$ can be calculated as
\be
\label{eq:T-fromT+}
T_-(\vartheta) = T_+(\vartheta) + 4 \, \int_{\vartheta_\mr{min}}^\vartheta \mr d \theta \, \frac{\theta}{\vartheta^2} T_+(\theta) \left[ 1 - 3 \left(\frac{ \theta}{\vartheta}\right)^2 \right]\,.
\ee
Conversely, one can construct $T_+$ for a given $T_-$.\\
The expressions for $T_+$ and $T_-$ used in this paper are given in the Appendix. We calculate the $EB$-statistics according to Eq. (\ref{eq:ebgeneral}) and compare the results to the ring statistics. Note that this $EB$-statistics can be optimized, e.g., with respect to its S/N ratio or its ability to constrain cosmology. For more details on this topic the reader is referred to \cite{fki09}.\\
In this paper, the $EB$-statistics is calculated as a function of $\Psi$. Similar to the ring statistics, $\Psi$ denotes the maximum angular scale of 2PCFs which enter in the calculation of $E (\Psi)$. 

\section{Covariance and signal-to-noise ratio}
\label{sec:covring}
For our further analysis we have to derive a formula to calculate the covariance of ring statistics and $EB$-statistics. A corresponding expression for $\map$ reads \citep[see e.g.][]{svm02}.
\beq
\label{eq:covmap}
\mr C_\mathcal M (\theta_k,\theta_l)) &=& \frac{1}{4}\sum^{I}_{i=1} \sum^{J}_{j=1} \frac{\Delta \vartheta_i \Delta \vartheta_j}{\theta^2_k \theta^2_l} \; \vartheta_i \vartheta_j \nn \\
&\times& \left[ \sum_{m,n=+,-}T_m \left( \frac{\vartheta_i}{\theta_k} \right) \; T_n \left( \frac{\vartheta_j}{\theta_l} \right) \mr C_{mn}(\vartheta_i,\vartheta_j) \right] \;,
\eeq
with $\mr C_{mn}(\vartheta_i,\vartheta_j)$ denoting the 2PCF covariance. Here, the upper limits $I$ and $J$ are chosen such that $\vartheta_i \leq 2 \theta_k$ and $\vartheta_j \leq 2 \theta_l$. The ring statistics' covariance is defined as
\be 
\label{eq:ringcov}
\mr C_{\mathcal R}(\Psi_k,\Psi_l)=\ensav{\hat R^2_\mr E (\Psi_k)\, \hat R^2_\mr E (\Psi_l)} - \mathcal \rrE (\Psi_k) \rrE(\Psi_l)\,,
\ee
where $\hat R^2_\mr E$ denotes the estimator of the ring statistics. To calculate this estimator from a binned 2PCF data vector with bin width $\Delta \vartheta_i$ we replace the integrals in Eq. (\ref{eq:ringe}) by a sum over the bins
\be 
\label{eq:ringesti}
\hat R^2_\mr E (\Psi)=\frac{1}{2} \sum_{i=1}^I \frac{\Delta \vartheta_i}{\vartheta_i} \left[ \hat \xi_+(\vartheta_i) \, Z_+(\vartheta_i,\eta) + \hat \xi_-(\vartheta_i) \, Z_-(\vartheta_i,\eta) \right] \,,
\ee 
with $\hat \xi_\pm (\vartheta_i)$ denoting the estimator of the $i$-th 2PCF bin. The upper limit $I$ in Eq. (\ref{eq:ringesti}) denotes the bin up to which $\vartheta_i \leq \Psi$. Inserting Eq. (\ref{eq:ringesti}) into Eq. (\ref{eq:ringcov}) we derive
\beq
\label{eq:cov_ring}
\mr C_{\mathcal R}(\Psi_k, \Psi_l) &=& \sum^{I}_{i=1} \sum^{J}_{j=1} \frac{\Delta \vartheta_i \Delta \vartheta_j}{4\,\vartheta_i \vartheta_j} \nn \\
&\times& \left[ \sum_{m,n=+,-}Z_m ( \vartheta_i ,\Psi_k) \; Z_n ( \vartheta_j,\Psi_l ) \, \mr C_{mn}(\vartheta_i,\vartheta_j) \right],
\eeq
where $I$ and $J$ denote the bins up to which $\vartheta_i \leq \Psi_k$ ($\vartheta_j \leq \Psi_l$) holds.\\
Similarly a covariance for the general $EB$-statistics can be calculated as
\beq
\label{eq:cov_eb}
\mr C_E (\Psi_k, \Psi_l) &=& \sum^{I}_{i=1} \sum^{J}_{j=1} \Delta \vartheta_i \Delta \vartheta_j \,\vartheta_i \vartheta_j \nn \\
&\times& \left[ \sum_{m,n=+,-}T_m \left( \vartheta_i ,\theta_k \right) \; T_n \left( \vartheta_j, \theta_l \right) \mr C_{mn}(\vartheta_i,\vartheta_j) \right] \,.
\eeq

\subsection{Correlation matrices}
\label{sec:correlation}
\begin{figure*}
    \includegraphics[width=17cm]{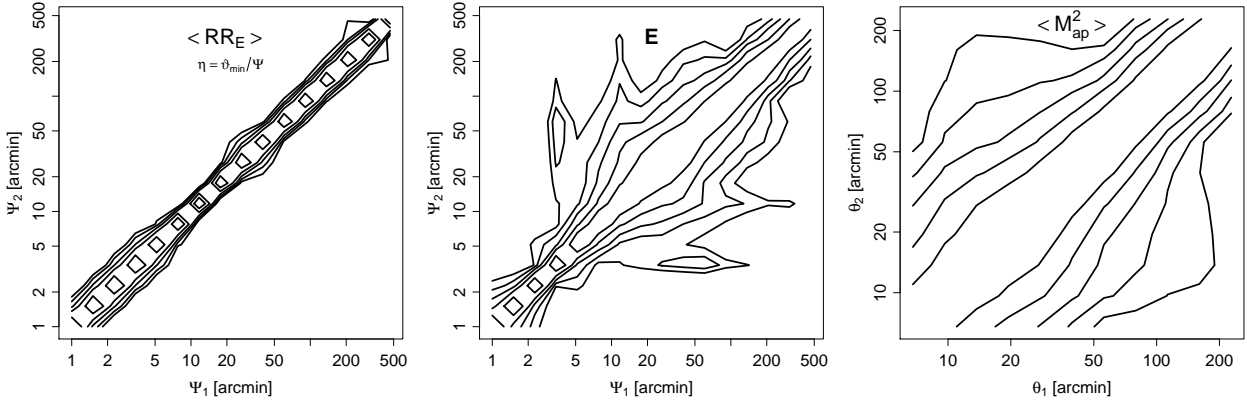}
  \caption{This figure shows the correlation matrices of $\rrE$ (\textit{left}), $E$ (\textit{middle}), and $\map$ (\textit{right}) derived from ray-tracing 2PCF covariance matrix. In each panel the $n$-th contour line (starting with $n=1$ close to the diagonal) marks values of $(0.8)^n$.}
         \label{fig:correlation_ringray_CFHTLS}
\end{figure*}

In order to illustrate the correlation between the individual data points we calculate the correlation matrix $\mathbf R$ for $\rrE$, $E$, and $\map$ from the corresponding covariance matrix. For $\C$ being the covariance of either $\rrE$, $E$, or $\map$ the correlation coefficients are defined as   
\be
\label{eq:rdef}
\mr R_{ij}=\frac{\mr C_{ij}}{\sqrt{\mr C_{ii} \mr C_{jj}}} \,.
\ee
The covariances are calculated from a 2PCF ray-tracing covariance via Eqs. (\ref{eq:covmap}), (\ref{eq:cov_ring}), and (\ref{eq:cov_eb}), respectively; finally the correlation matrix is obtained via Eq. (\ref{eq:rdef}). The ray-tracing simulations (175 realizations) have the following underlying cosmology: $\om=0.27, \Omega_\Lambda=0.73, \sig=0.78, h=0.73, \Omega_\mr b=0.044, n_\mr s=1.0$. From now on we refer to this cosmological parameter set as our fiducial cosmological model $\vec \pi_\mr{fid}$. Survey parameters which enter in the calculation read as follows: galaxy density $n_\mr{gal}=25/\mr{arcmin}^2$, survey area $A=36$ $\mr{deg}^2$, and intrinsic ellipticity noise $\sigma_\epsilon=0.38 $. The survey parameters differ slightly from those of the covariance used in the latest CFHTLS survey; FSH08 use $A=34.2$ $\mr{deg}^2$, $n_\mr{gal}=13.3/\mr{arcmin}^2$, and $\sigma_\epsilon=0.42 $.\\
The covariance matrices have a different angular range corresponding to the data vectors of $\rrE$, $E$, and $\map$, which we define as
\beq
\label{eq:vectorring1}
\vec{\rrE} &=& \left[ \rrE (\Psi_1), ..., \rrE (\Psi_n) \right]^{\mr t} \,,\\
\label{eq:vectorring2}
\vec{E} &=& \left[ E (\Psi_1), ..., E (\Psi_n) \right]^{\mr t} \,, \\
\label{eq:vectorring3}
\vec{\langle M_\mr{ap}^2\rangle} &=&\left[ \map (\theta_1),.., \map (\theta_m)\right]^{\mr t} \,.
\eeq
Whereas $\vec \rrE$ and $\vec E$ extend from $1' \leq \Psi \leq 460'$, $\vec \map$ extends from $6' \leq \theta \leq  230'$. The different maximum angular separation of the aperture mass dispersion result from the fact that $\rrE (\Psi)$ and $E (\Psi)$ contain information on the 2PCF with $\vartheta \leq \Psi$, whereas $\map (\theta)$ contains information on the 2PCF up to $\vartheta \leq 2 \theta$. The lower limit of $6'$ was chosen to circumvent the problem of E/B-mode mixing for the $\map$ covariance. The range of the original 2PCF ray-tracing covariance extends from $0\farcm5 \leq \vartheta \leq 460'$. Below $6'$ it is not possible to calculate the $\map$ covariance properly.\\
Figure \ref{fig:correlation_ringray_CFHTLS} shows the correlation matrices of the ring statistics (left), the $EB$-statistics (middle), and of the aperture mass dispersion (right). Starting from the diagonal, where $R_{ii}=1$, the $n$-th contour line corresponds to values of $0.8^n$. It is clearly noticeable that data points of the ring statistics are significantly less correlated than those of the aperture mass dispersion and the $EB$-statistics. \\
The boxy contours in Fig. \ref{fig:correlation_ringray_CFHTLS} result from the small number of bins we choose in the covariances. The reason for this is that the ray-tracing covariance is an estimated quantity; its inverse, needed for  the likelihood analysis in Sect. \ref{sec:mapvsring}, is in general affected from numerical artifacts. These artifacts become more severe in case of a high dimension matrix. In order to guarantee a stable inversion process we choose a small number of bins.
\subsection{Signal-to-noise ratio}
\label{sec:signoise}
\begin{figure*}
\sidecaption
 \includegraphics[width=13cm]{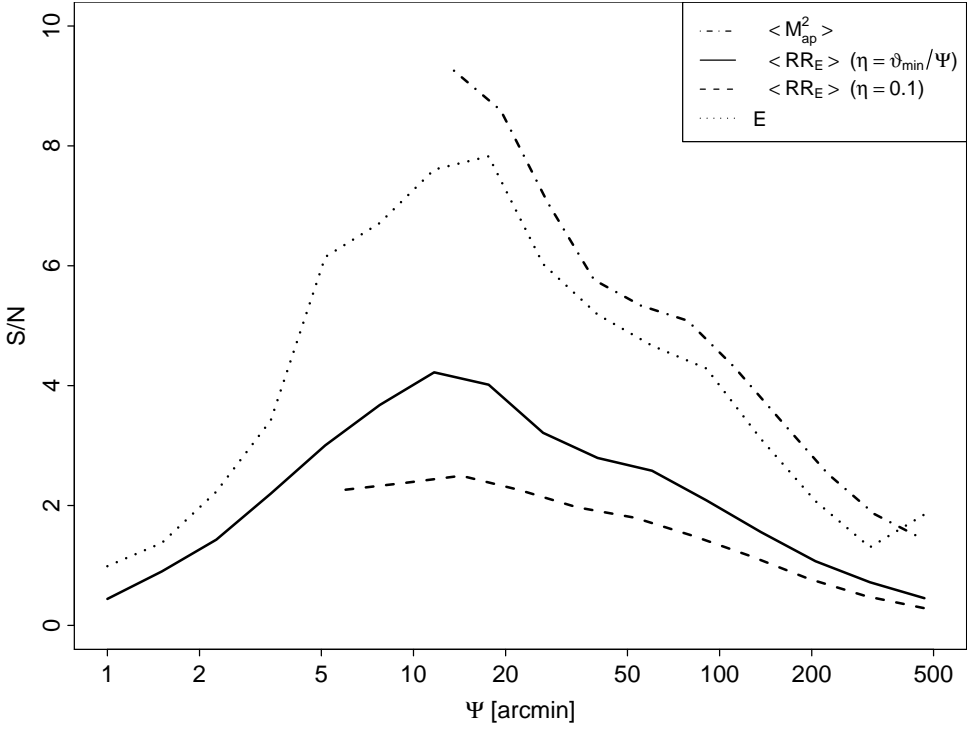}
   \caption{The S/N ratio of the ring statistics (for $\eta=0\farcm5/\Psi$ and $\eta=0.1$), the $EB$-statistics, and the aperture mass dispersion calculated from a set of theoretical 2PCFs with $\vartheta \in [0\farcm5;460']$. The different angular range of the measures is explained in the text. }
    \label{fig:signoise}
\end{figure*}
We now use the above derived covariances to quantify the S/N ratio of the ring statistics, $EB$-statistics and compare both to that of the aperture mass dispersion.\\
We calculate a set of 2PCFs via Eq. (\ref{eq:xi+-}) for an angular range similar to that of the ray-tracing simulations (see Sect. \ref{sec:correlation}), i.e. $\vartheta \in [0\farcm5,460']$. The required shear power spectra $P_\mr E$ are obtained from the density power spectra $P_\delta$ employing Limber's equation. As underlying cosmology we choose our fiducial model (see Sect. \ref{sec:covring}). The power spectrum $P_\delta$ is calculated from an initial Harrison-Zeldovich power spectrum ($P_{\delta}(k) \propto k^{n_\mr s}$) with the transfer function from \cite{ebw92}. For the non-linear evolution we use the fitting formula of \cite{sm03}. In the calculation of $P_\mr E$ we choose a redshift distribution of source galaxies similar to that of \cite{bhs07}
\be 
\label{eq:redshiftben}
n(z)=\frac{\beta}{z_0 \Gamma \left( \left(1+\alpha \right)/\beta \right)} \left( \frac{z}{z_0}\right)^\alpha \exp \left[ - \left(  \frac{z}{z_0} \right)^\beta \right]\,,
\ee 
with $\alpha=0.836$, $\beta=3.425$, $z_0=1.171$. \\
From this set of 2PCFs we calculate data vectors of $\rrE$, $E$, and $\map$ according to Eqs. (\ref{eq:ringe}) and (\ref{eq:ebgeneral}), and (\ref{eq:mapfromxi}), respectively.
The angular range of these data vectors are chosen similar to the range of the corresponding covariances (Sect. \ref{sec:correlation}), i.e. $0\farcm5 \leq \Psi \leq 460\farcm0$ for $\vec{\rrE}$ and $\vec E$, and $6\farcm0 \leq \theta \leq  230\farcm0$ for $\vec{\langle M_\mr{ap}^2\rangle}$. The S/N ratio is calculated as
\be
\label{eq:signoise}
\mr{S/N}=\frac{\rrE(\Psi_i)}{\left[\mr C_{\mathcal R}(\Psi_i,\Psi_i)\right]^{1/2}} \quad \mr{and} \quad \mr{S/N}=\frac{\map(\theta_i)}{\left[\mr C_{\mathcal M}(\theta_i,\theta_i)\right]^{1/2}} \,.
\ee
The results are illustrated in Fig. \ref{fig:signoise}. We compare the ring statistics for with scale-dependent $\eta$ and $\eta=0.1$ to the $EB$-statistics and the aperture mass dispersion. The figure shows the anticipated behavior (Sect. \ref{sec:ringstatistics}); the ring statistics with scale-dependent $\eta$ gives a larger S/N ratio compared to the case where $\eta$ is fixed. In addition, it can be measured down to arbitrary small values of $\Psi$ (above $\vartheta_\mr{min}$), which is not possible when choosing a fixed $\eta$. For the case considered here, i.e. $\vartheta_\mr{min}=0\farcm5$, the choice of $\eta=0.1$ already limits the range of $\Psi$ to scales $\geq 5'$; decreasing $\eta$ further in order to increase the S/N ratio will limit $\vec{\rrE}$ to larger $\Psi$. \\
When comparing the ring statistics to the aperture mass dispersion, we find that the ring statistics' signal is lower. Even with the scale-dependent filter function the S/N ratio of the ring statistics is on average by a factor of $\approx 2$ smaller than the S/N ratio of the aperture mass dispersion. This difference can be explained when comparing the filter functions of $\rrE$ and $\map$, $Z_\pm$ (Fig. \ref{fig:zfunc_CFHTLS}) and $T_\pm$ \citep[e.g. Fig. 1 in][]{svm02}, respectively. The $Z$-functions have two roots at their boundaries whereas the $T_+$-function becomes particularly large for small $x$. However, we point out that the S/N ratio does not solely determine the ability of a measure to constrain cosmology, but one has to account for the fact that the data points of the $\rrE$ are less correlated than those of $\map$. For a full comparison of the information content we examine both measures in a likelihood analysis.\\
Compared to the ring statistics the S/N ratio of the $EB$-statistics is significantly larger on all scales, which again can be explained by the fact that the filter function of the $EB$-statistics does not have roots at their boundaries. Compared to the aperture mass dispersion, the $EB$-statistics' S/N ratio is slightly lower. However, we point out that the $EB$-filter function, we chose here, is a simple second-order polynomial. We will present an extended analysis of this general $EB$-filter functions in a future paper.
\section{Comparison of the information content of $\rrE$ and $\map$}
\label{sec:mapvsring}
\begin{figure*}
\includegraphics[width=6cm]{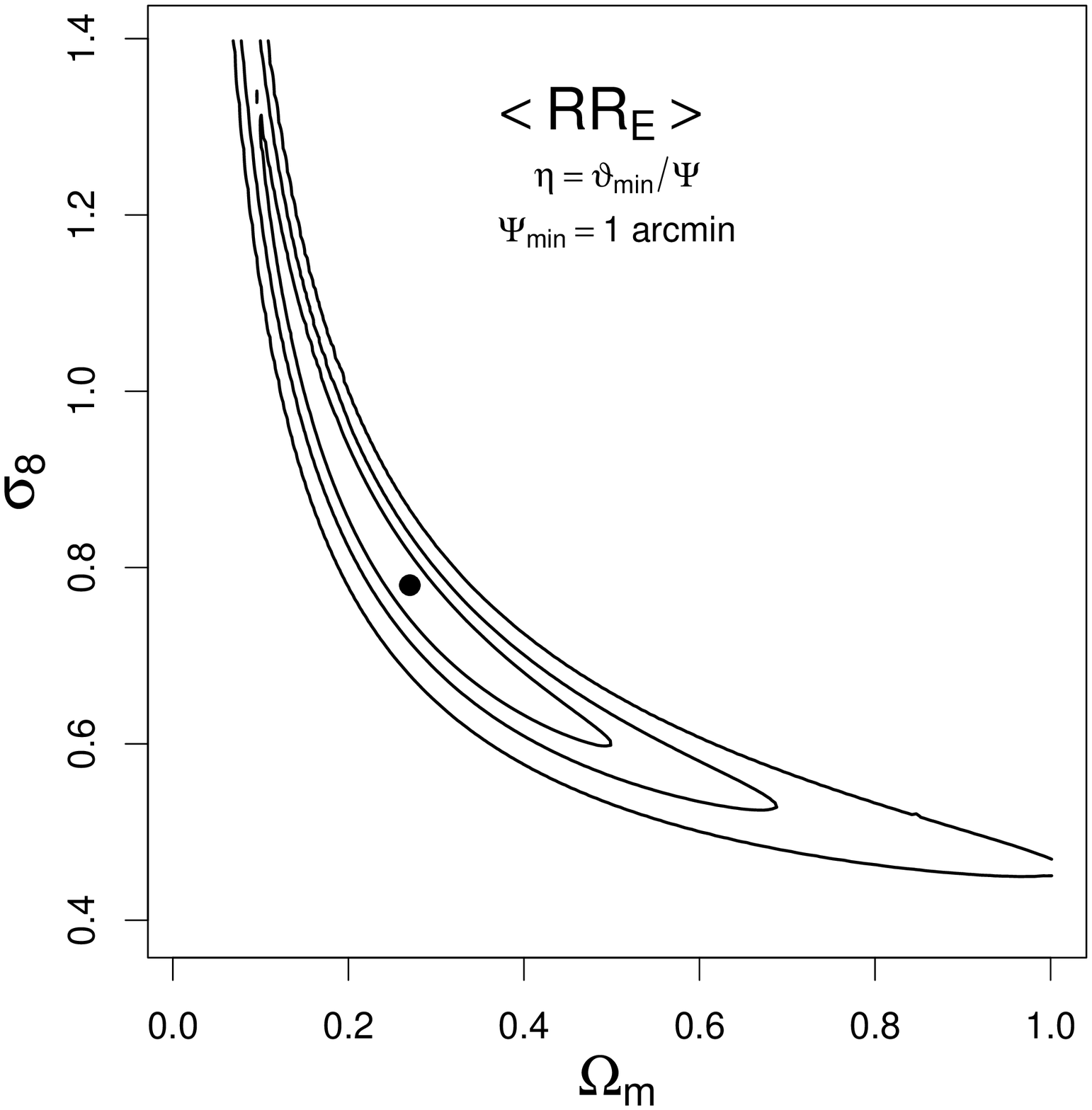}
\includegraphics[width=6cm]{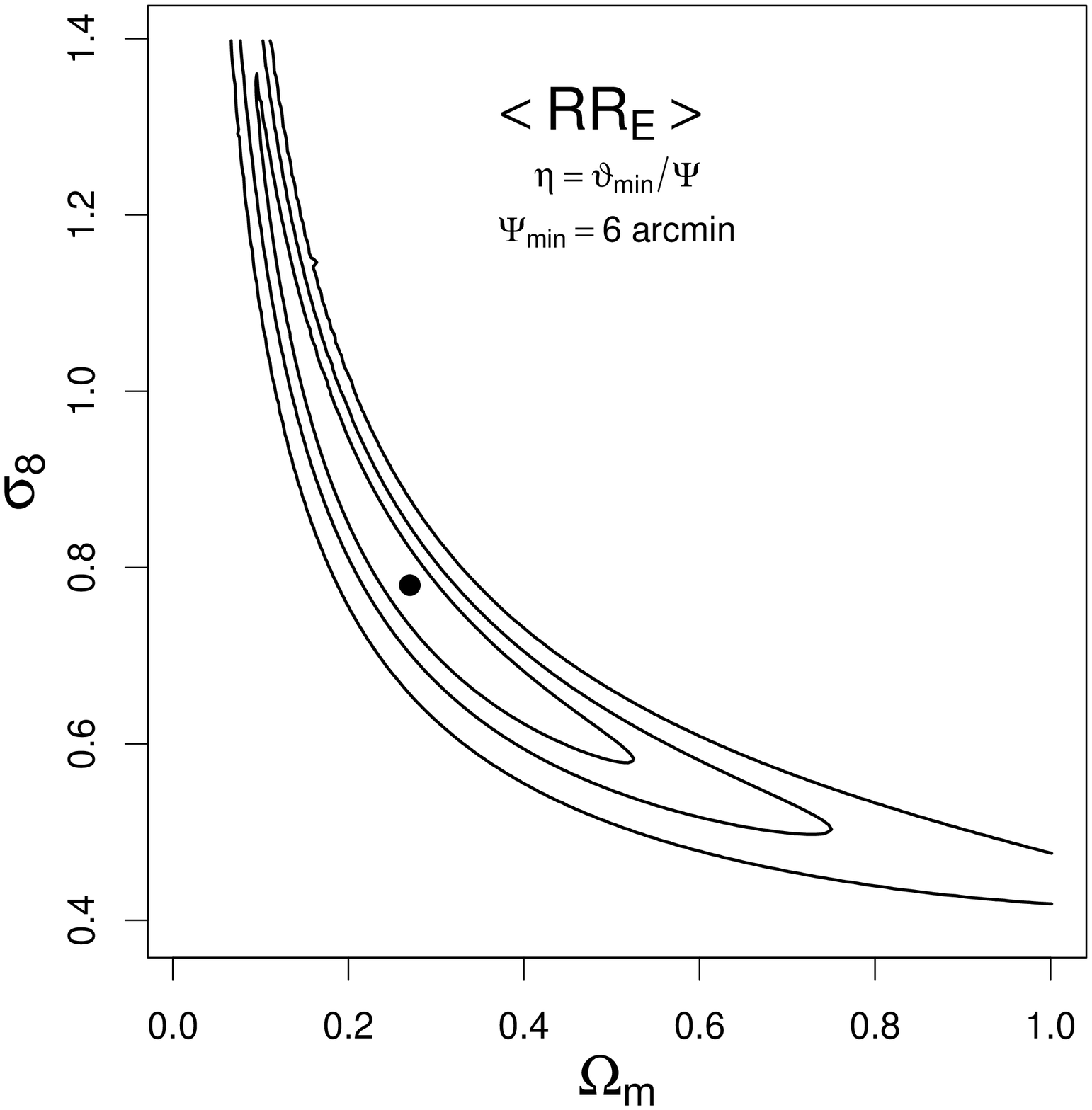}
\includegraphics[width=6cm]{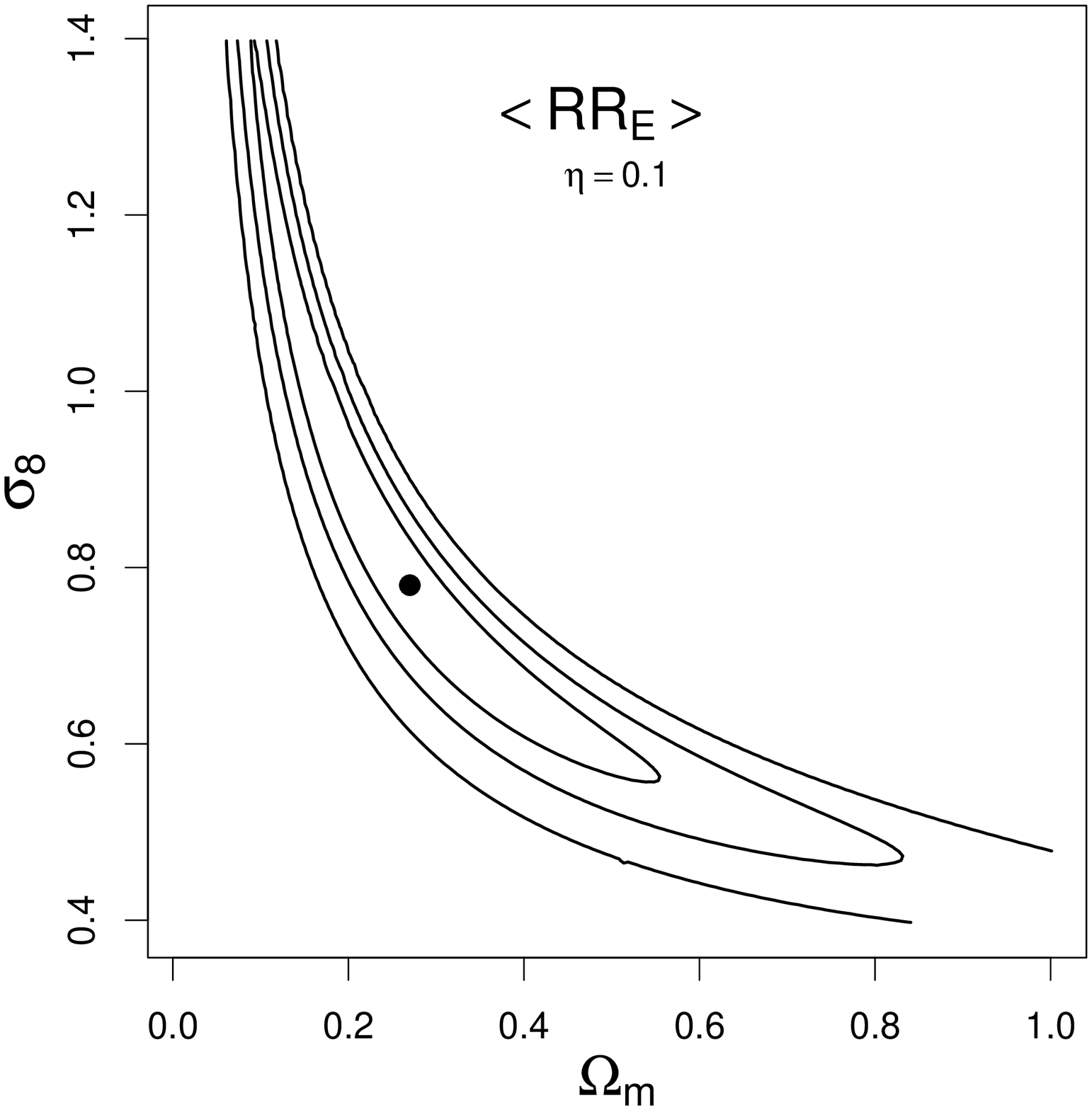}
\sidecaption
\includegraphics[width=6cm]{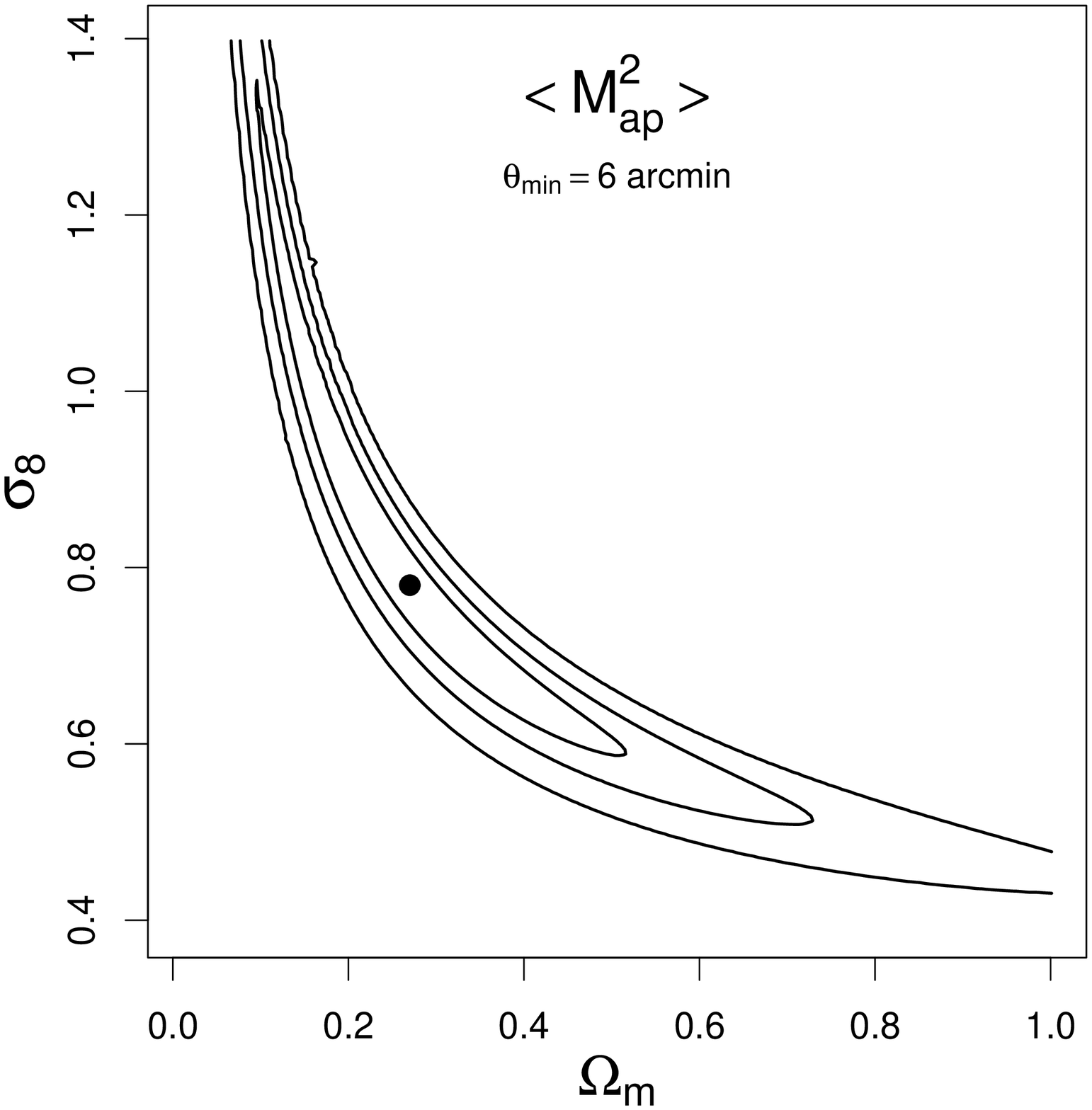}
\includegraphics[width=6cm]{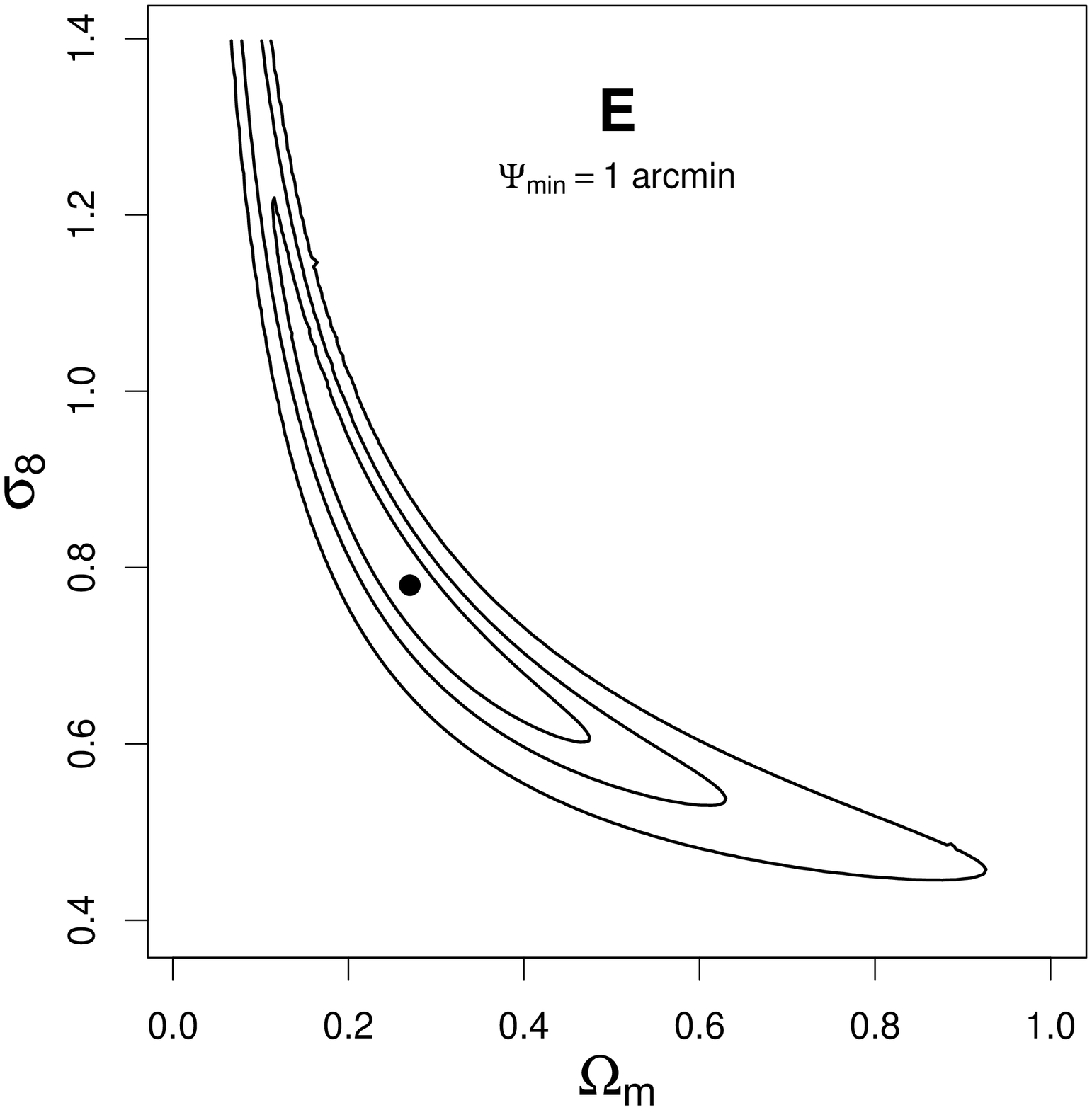}
  \caption{The $68\%$-, $95\%$-, $99.9\%$-contours of the likelihood analysis using the ring statistics, the $EB$-statistics, and the aperture mass dispersion. We compare 5 different cases, namely in the \textit{upper} row: $\eta=\Psi/\vartheta_\mr{min}$ for $\Psi_\mr{min}=1'$ (\textit{left}), and for $\Psi_\mr{min}=6'$ (\textit{middle}), and $\rrE$ with $\eta=0.1$ (\textit{right}). In the \textit{lower} row we see $\map$ (\textit{left}), and the $EB$-statistics for $\Psi_\mr{min}=1'$ (\textit{right}). The data vectors are calculated analytically from a power spectrum; the covariance is obtained from ray-tracing simulations. The filled circle marks our fiducial cosmology.}
         \label{fig:ring_info}
\end{figure*}

We now perform a likelihood analysis in the $\om$ vs. $\sig$ parameter space in order to compare the ability of $\rrE$, $E$, and $\map$ to constrain cosmological parameters. We calculate 2PCF data vectors for various combinations of $\sig \in [0.4;1.4]$ and $\om \in [0.01;1.0]$, therefrom derive the data vectors of $\rrE$, $E$, and $\map$ and test these against the corresponding data vectors obtained from our fiducial model (Sect. \ref{sec:correlation}). We assume that all data vectors are normally distributed in parameter space and calculate the posterior likelihood according to Bayes theorem. Our likelihood function $p(\vec d|\vpi)$ then reads 
\be
\label{eq:likefunc}
p(\vec d|\vpi) = \frac{\exp \, \left[ -\frac{1}{2} \, \left( (\vec d(\vpi) - \vec d (\vpi_\mr{fid}))^\mr t \;\C^{-1} \;( \vec d(\vpi) - \vec d (\vpi_\mr{fid}))\right) \right]}{(2 \pi)^{n/2} \; |\C|^{\frac{1}{2}}}  \,,
\ee
where $\vec d$ must be replaced by the considered data vector, either $\vec \rrE$ (Eq. \ref{eq:vectorring1})
, $\vec \map$ (Eq. \ref{eq:vectorring2}), or $\vec E$ (Eq. \ref{eq:vectorring3}).\\
To illustrate the information content we calculate the so-called credible regions, where the true parameter is located with a probability of 68\%, 95\%, 99,9\%, respectively. In addition, we quantify the size of these credible regions through the determinant of the second-order moment of the posterior likelihood \citep[see e.g.][]{eks08} 
\be
\label{eq:quadrups}
\mathcal Q_{ij} \equiv \int \tn d^2 \vpi p(\vpi|\vxi) \; (\pi_i-\pi_i^{\mr f})(\pi_j -\pi_j^{\mr f})\,,
\ee
where $\pi_i$ are the varied parameters, $\pi_i^{\mr f}$ are the parameter of the fiducial model ($i=1,2$, corresponding to $\om$ and $\sig$). The square root of the determinant is given by
\be
\label{eq:detquadrups}
q= \sqrt{|\mathcal Q_{ij}|} = \sqrt{\mathcal Q_{11} \mathcal Q_{22} - \mathcal Q_{12}^2},
\ee 
and it can be considered as our figure-of-merit quantity. Smaller credible regions in parameter space correspond to a smaller value of $q$. In this paper all $q$'s are given in units of $10^{-4}$.\\ 
For the likelihood analysis in this section we employ the ray-tracing covariances and choose the angular range of the data vectors correspondingly (Sect. \ref{sec:correlation}), i.e. $\Psi \in [1';460']$ and $\theta \in [6';230']$. We further assume a flat prior probability with cutoffs, which means $p(\vpi)$ is constant for all parameters inside a fixed interval ($\om \in [0.01;1.0]$, $\sig \in [0.4;1.4]$) and $p(\vpi)=0$ else.\\
As we obtain our covariance from ray-tracing simulations we automatically account for the non-Gaussianity of the shear field, however we neglect the cosmology dependence of the covariance \citep[for more details see][]{esh08}. Furthermore, we account for the bias which occurs during the inversion of the ray-tracing covariance by applying the correction factor outlined in \cite{har07}.\\ 
The upper row of Fig. \ref{fig:ring_info} shows the result of the likelihood analysis for the ring statistics. We consider 3 cases: First, $\rrE$ with $\eta=\vartheta_\mr{min}/\Psi$ and $\Psi \in [1';460']$ (left). Second, $\rrE$ with $\eta=\vartheta_\mr{min}/\Psi$ and $\Psi \in [6';460']$ (middle). Third, $\rrE$ with $\eta=0.1$ and $\Psi \in [1';460']$ (right). The lower row shows a similar analysis for $\map$ with an angular range $\theta \in [6';230']$ (left) and the $EB$-statistics for $\Psi \in [1';460']$ (right). The black, filled circle indicates the fiducial cosmology; the contours correspond to the aforementioned credible regions. In addition we quantify the information content by the values of $q$, defined in Eq. (\ref{eq:detquadrups}), which are summarized in Table \ref{tab:qvalues}. \\
The ring statistics with $\eta=\vartheta_\mr{min}/\Psi$ is a clear improvement over $\rrE$ with $\eta=0.1$ which can be explained by the larger S/N ratio of the first compared to the second. Considering the ring statistics with scale-dependent $\eta$, we find that adding information below $6'$ increases the information content of $\rrE$, such that it gives tighter constraints than the $\map$ data vector. The strength of this gain in information can be explained by the small correlation of ring statistics' data points.\\
In our analysis it was not possible to calculate $\map$ for $\theta \leq 6'$ due to the aforementioned E/B-mode mixing, however this can change if the 2PCF is measured on smaller angular scales. For this case we expect the improvement of ring statistics over the aperture mass dispersion to be even more significant. Due to the lower correlation of the ring statistics' data points an inclusion of smaller scales will enhance constraints from $\rrE$ stronger than those from $\map$.\\ 
The $EB$-statistics gives tighter constraints on cosmology than the optimized ring statistics, which can be explained by its larger S/N ratio. However, we do not use the $EB$-statistics to analyze the CFHTLS data in the next section for the reason that the $EB$-statistics' data points are strongly correlated (see Fig. \ref{fig:correlation_ringray_CFHTLS}). In order to identify B-modes as a function of angular scale accurately, the lower correlation of the ring statistics is more useful.
\begin{table}
\caption{Values of $q$ resulting from the likelihood analyses of the 5 data vectors.}
\centering
\label{tab:qvalues}
\begin{tabular}{l l }\hline \hline
 Data vector & $q$  \\ \hline
$\rrE$ ($\eta=\vartheta_\mr{min}/\Psi$, $\Psi_\mr{min}=1'$) & 153.8\\
$\rrE$ ($\eta=\vartheta_\mr{min}/\Psi$, $\Psi_\mr{min}=6'$) & 177.3\\
$\rrE$ ($\eta=0.1$)& 207.9\\
$\map$ ($\theta_\mr{min}=6'$)& 169.8\\
$E$ ($\Psi_\mr{min}=1'$)& 122.5\\ \hline
\end{tabular}
\end{table}
\section{Ring statistics with the CFHTLS} 
\label{sec:cfhtlsring}
\begin{figure*}
   \includegraphics[width=6cm]{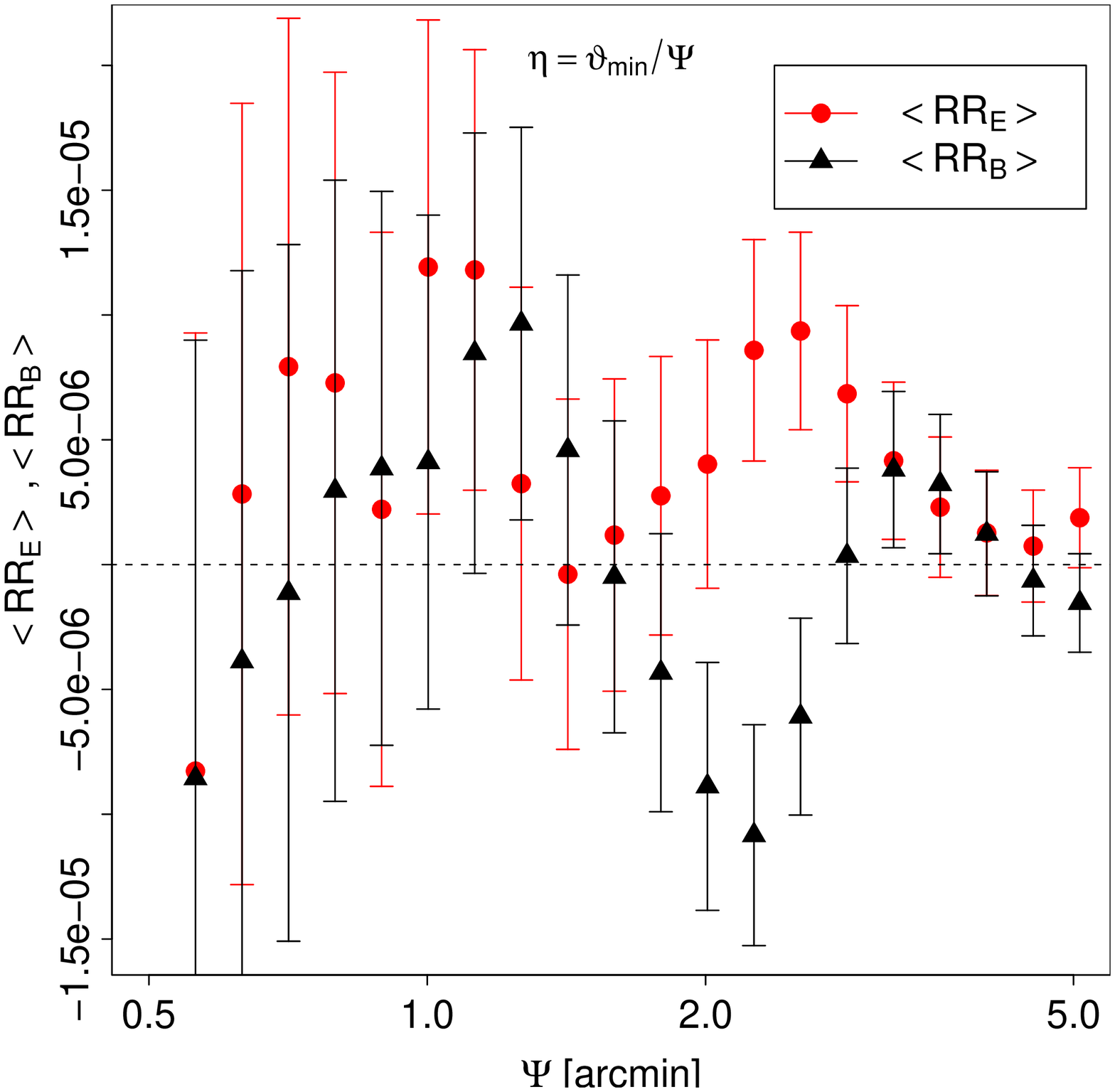}
  \includegraphics[width=6cm]{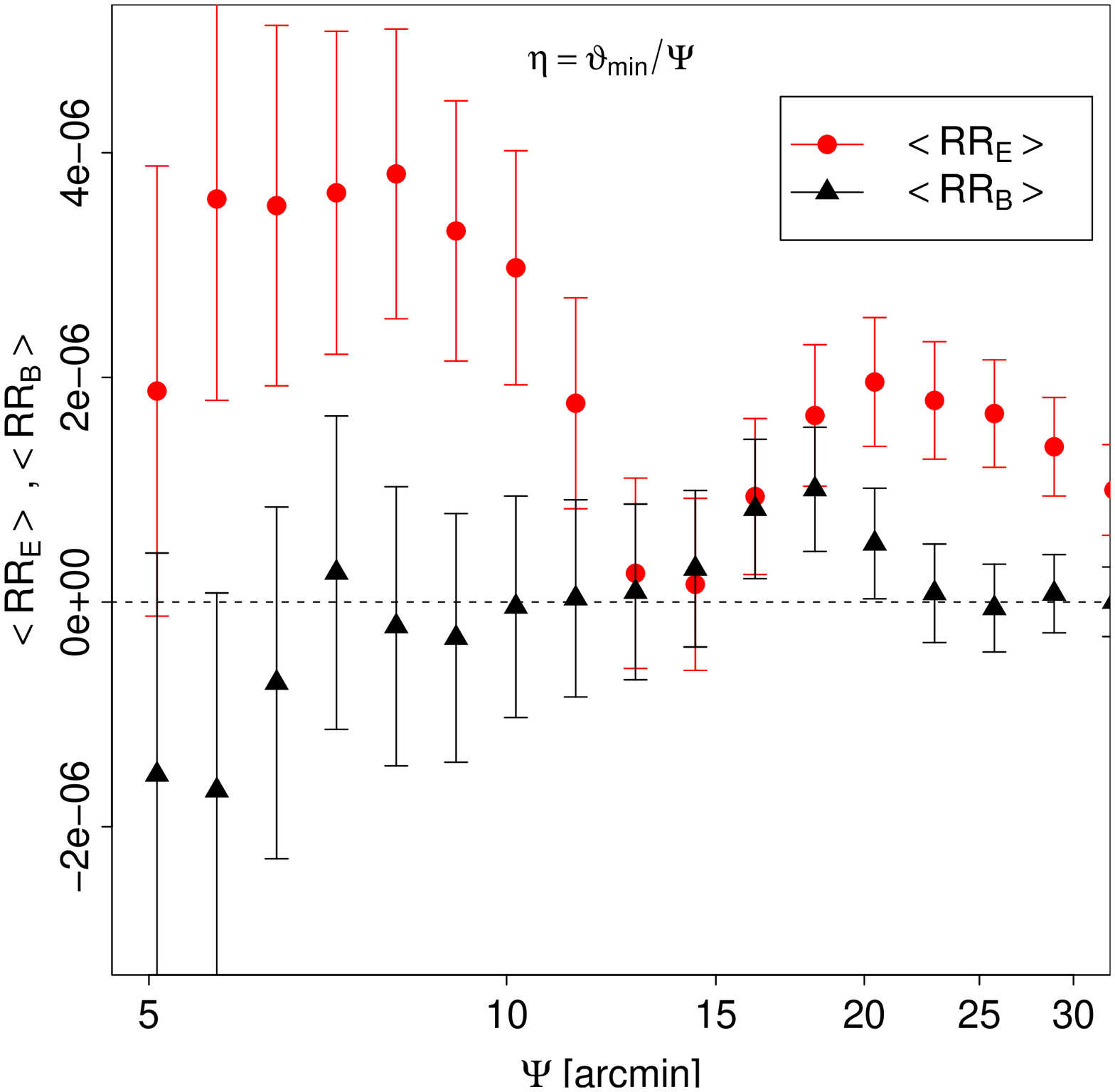}
  \includegraphics[width=6cm]{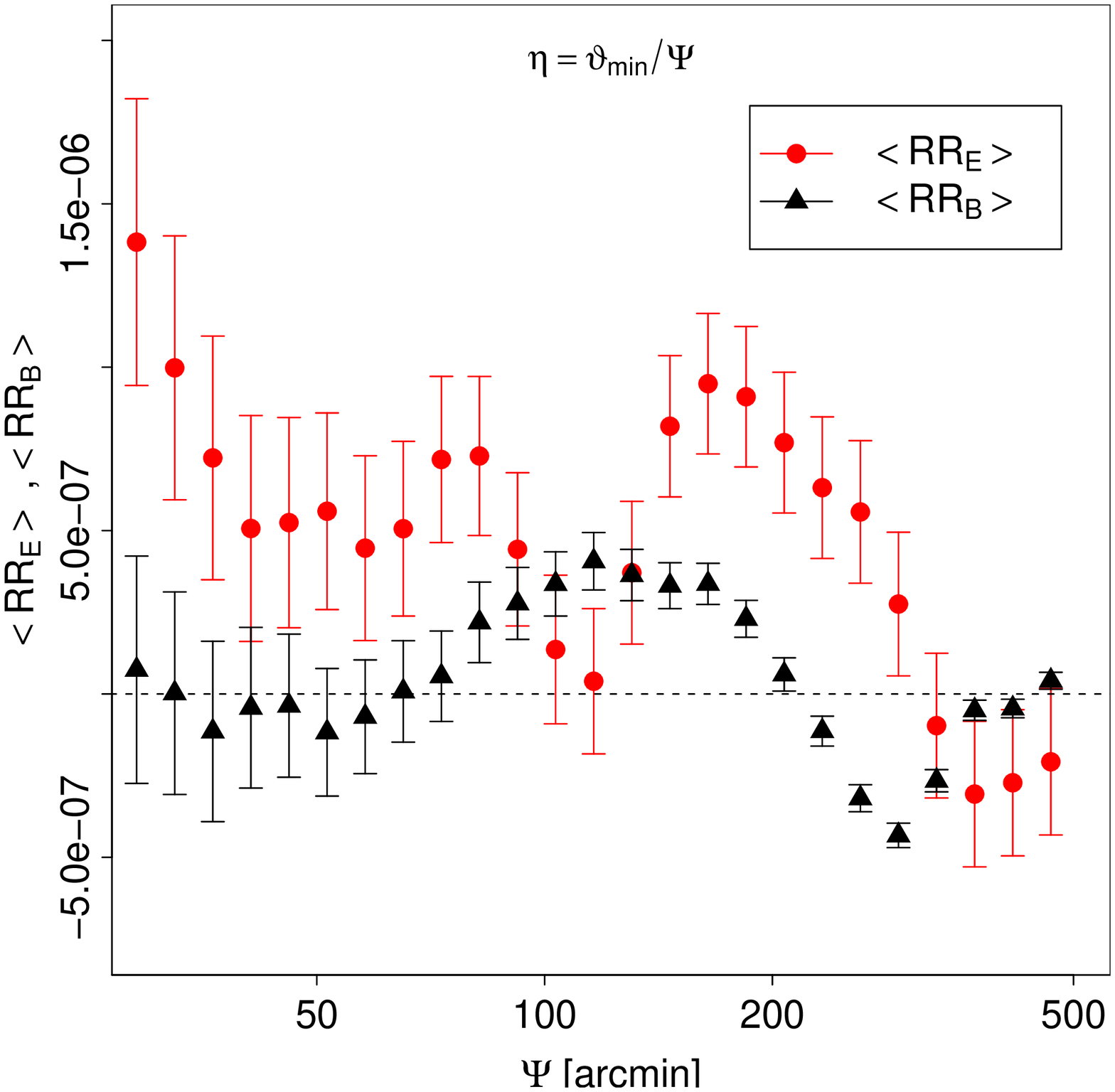}
   \includegraphics[width=6cm]{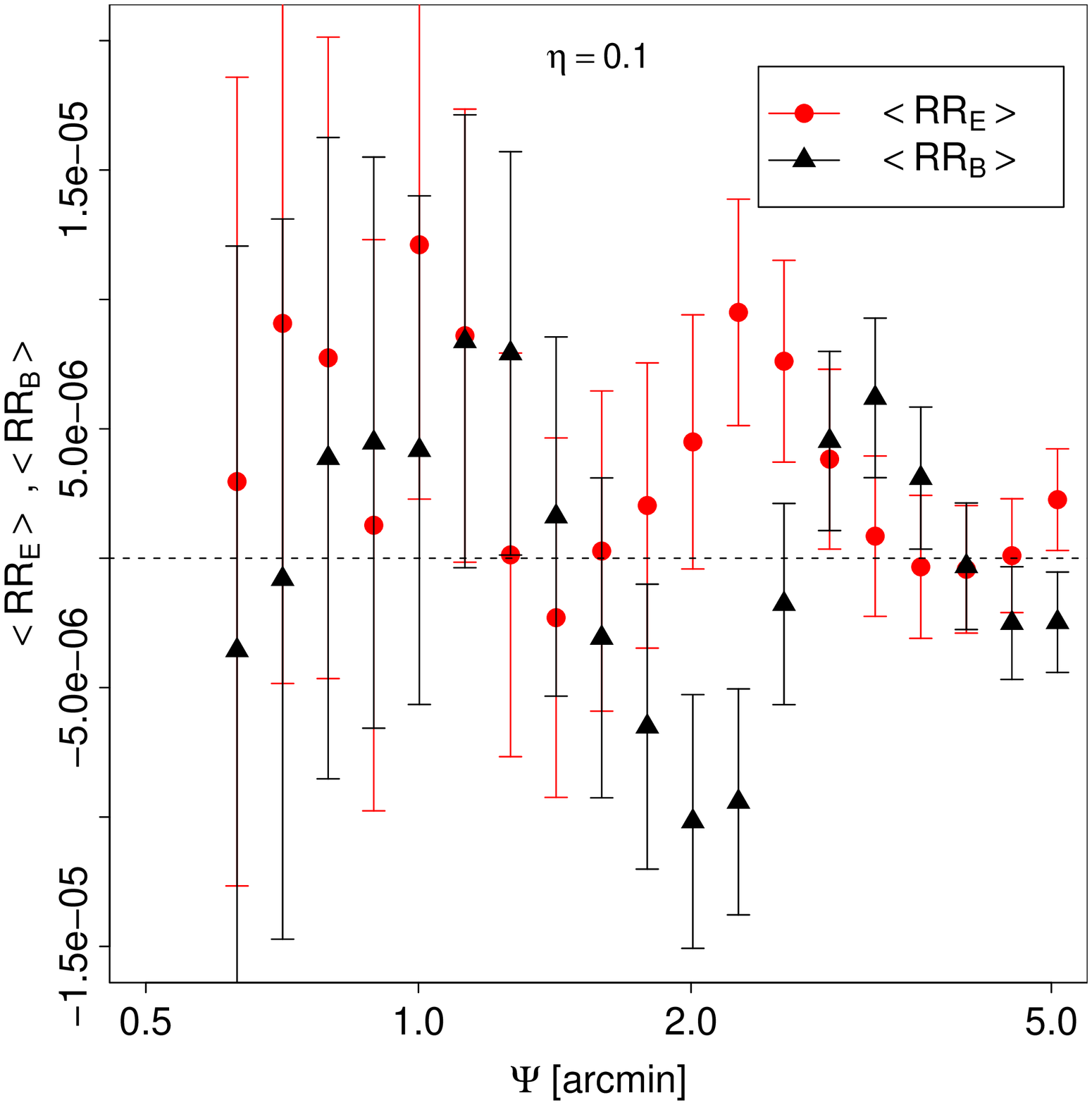}
  \includegraphics[width=6cm]{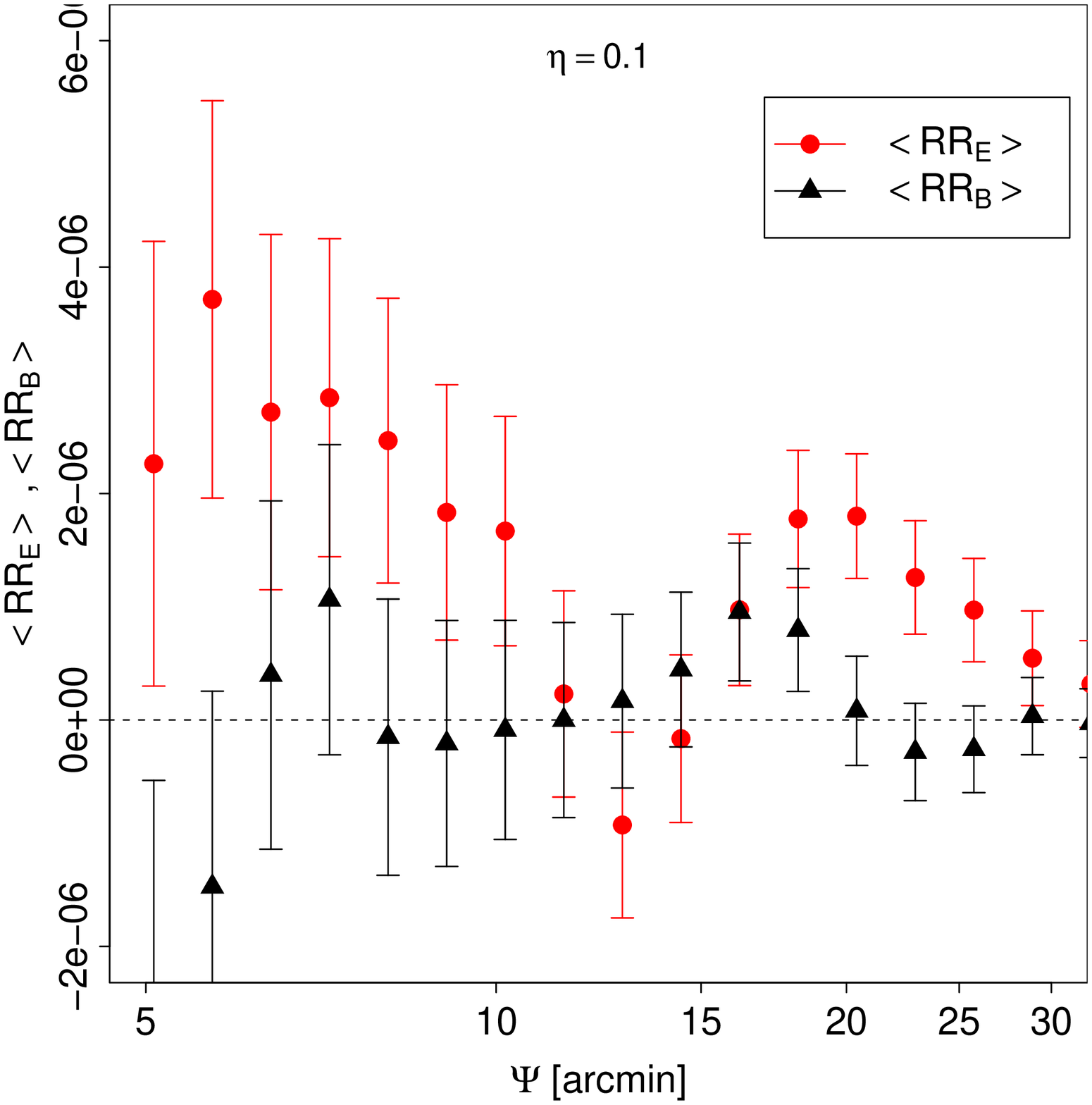}
  \includegraphics[width=6cm]{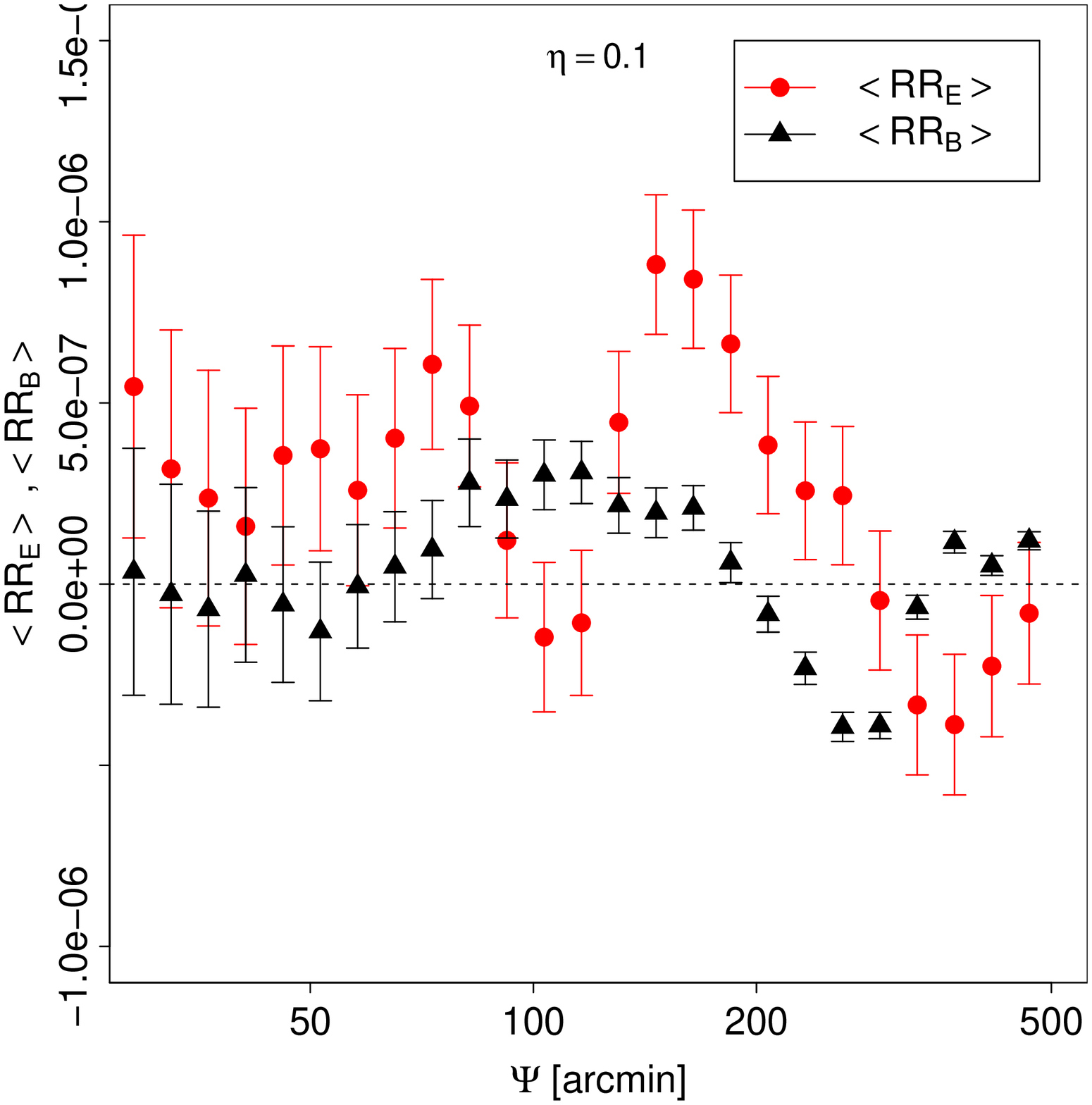}
   \caption{The ring statistics signal measured from the CFHTLS for the case of $\eta=\vartheta_\mr{min}/\Psi$ (\textit{upper} row). The red data points (circles) correspond to the E-mode signal, the black data points (triangles) to the B-mode signal. The three panels correspond to small (\textit{left}), intermediate (\textit{middle}), and large (\textit{right}) scales. The \textit{lower} row shows a similar analysis but for $\eta=0.1$. }
         \label{fig:ring_CFHTLS}
\end{figure*}

In Sect. \ref{sec:correlation} we have shown that the ring statistics' data points are significantly less correlated compared to data points of the aperture mass dispersion. Therefore, despite its lower S/N ratio, the ring statistics provides an ideal tool to analyse B-mode contaminations depending on the angular scale. In this section we use the 2PCFs of the FSH08 analysis and therefrom calculate the ring statistics for a scale-dependent $\eta=\vartheta_\mr{min}/\Psi$ and for $\eta=0.1$. We performed a similar analysis for other cases of fixed $\eta$, which resulted in a significantly weaker signal.\\ 
The CFHTLS 2PCF was measured in 72000 bins over an angular range of $0\farcm05 \leq \vartheta \leq 466'$; we calculate $\rrE$ (Eq. \ref{eq:ringe}) and $\rrB$ (Eq. \ref{eq:ringb}) in 60 logarithmic bins over a range $0\farcm5 \leq \Psi \leq 460\farcm0$. The error for the $i$-th E/B-mode data point is calculated as $\sqrt{\C_{R_{\mr E/\mr B}}(\Psi_i,\Psi_i)}$, where $\C_{R_{\mr E/\mr B}}(\Psi_i,\Psi_i) $ is calculated from a Gaussian 2PCF covariance. This Gaussian covariance was calculated from a theoretical model using the same cosmology and survey parameters as in the FSH08 analysis. We do not employ the non-Gaussian correction of \cite{svh07} as this corrects the $C_{++}$-term in the 2PCF covariance, but not the $C_{--}$- and $C_{+-}$-terms. Here, we use the full 2PCF covariance in the analysis. Similar to FSH08 we do not consider systematic errors in our analysis which might lead to an underestimation of the error bars.\\
The results of our analysis are illustrated in Fig. \ref{fig:ring_CFHTLS}. The three panels in the upper row show the ring statistics' E- and B-modes on (from left to right) small, intermediate and large scales of $\Psi$ for the case of $\eta=\vartheta_\mr{min}/\Psi$. The three panels in the lower row show the same analysis but for $\eta=0.1$. The circled (red) data points correspond to the E-mode signal, the triangled (black) data points correspond to the B-mode signal. \\
We measure a robust E-mode shear signal, however we also find a significant B-mode contribution on small (around $2'$), intermediate ($16'$-$22'$), and large scales (right panel). On small scales E-and B-mode are of similar order. It should be stressed that such an analysis of small-scale contaminations is not feasible with the aperture mass dispersion, which, to avoid the E/B-mode mixing on small scales, involves a theoretical (therefore B-mode free) 2PCF in its calculation. This theoretical data extension, combined with the fact that the aperture mass dispersion data points are stronger correlated (Sect. \ref{sec:covring}) can hide possible small-scale contaminations in the data.\\
The B-mode contamination on large scales is also observed in the FSH08 analysis. In addition, we find a small B-mode on intermediate scales (between $16'$ and $22'$), otherwise these intermediate scales are mostly free of B-modes and give a robust E-mode signal. The small correlation of the individual data points leads to the oscillations in the amplitude of the shear signal. A similar analysis with the aperture mass dispersion shows a much smoother behavior. 
\section{Conclusions}
\label{sec:conc}
Decomposing the shear field into E- and B-modes is an important check for systematics in a cosmic shear analysis. The most commonly used methods for E- and B-mode decomposition, namely the aperture mass dispersion and the E/B-mode shear correlation function, require the 2PCF to be known down to arbitrary small or up to arbitrary large angular separations. In practice, the 2PCF is only measured over a finite interval $[\vartheta_\mr{min};\vartheta_\mr{max}]$. As a result the aforementioned methods do not separate E- and B-modes properly, e.g. the aperture mass dispersion suffers from E/B-mode mixing on small angular scales (see KSE06 for further details).\\
In contrast, the ring statistics (invented in SK07) separates E- and B-modes properly using 2PCFs measured on a finite angular scale. As outlined in SK07 the filter functions of the ring statistics, i.e. $Z_\pm$, are in general complicated to calculate; the authors restrict the free parameters this filter function to one, namely $\eta$. This parameter is held fixed, independent of the angular scale $\Psi$ at which the ring statistics is evaluated. In this paper, we improve on the condition of a fixed $\eta$ by choosing a scale-dependent $\eta=\vartheta_\mr{min}/\Psi$ which significantly improves on the ring statistics' S/N ratio. \\
Furthermore, we present a formula to calculate the ring statistics' covariance from a 2PCF covariance. This formula is applied to a 2PCF covariance obtained from ray-tracing simulations. We therefrom calculate the correlation matrices of ring statistics and aperture mass dispersion and find that the data points of the first are significantly less correlated than the data points of the second. We employ these covariances to compare the information content of the two second-order statistics and find that the ring statistics' data points place tighter constraints on cosmological parameters than data points of the aperture mass dispersion. The reason for this is that we can include smaller scales in the ring statistics's data vector which is not possible for $\map$ due to the aforementioned E/B-mode mixing. In addition, we consider a polynomial filter function which decomposes E- and B-modes on a finite interval and therefrom calculate an additional second-order measure, the $EB$-statistics. We compare the correlation of data points and the information content of this $EB$-statistics to the ring statistics and find that it shows a significantly larger correlation of the data points, but a higher information content. This can be explained by the high S/N ratio of the $EB$-statistics.\\
We apply the ring statistics with $\eta=\vartheta_\mr{min}/\Psi$ and $\eta=0.1$ to CFHTLS data, more precisely we calculate both from the 2PCF used in the latest CFHTLS analysis (FSH08). We measure a clear shear signal for $\eta=\vartheta_\mr{min}/\Psi$ which decreases when performing the same analysis for $\eta=0.1$. The fact, that data points of the ring statistics have small correlations enables us to determine the contaminated scales very accurately. We find B-modes on large scales which is comparable to the findings of FSH08. In addition, we detect B-modes on intermediate ($16'$ - $22'$) scales and a scattered B-mode contribution on scales below $3'$. In the latter case the shear signal is of the same order as the B-mode contribution.\\
A similar analysis with the aperture mass dispersion is only possible when including a 2PCF from a theoretical model in order to avoid the E/B-mode mixing on small angular scales. These added theoretical data can conceal remaining systematics (B-modes) which can be identified properly using the ring statistics. This property is most likely the most useful feature of the ring statistics. It can be used to detect remaining systemics very accurately in future surveys.\\ 
The noise-level of the ring statistics on small scales can be reduced by increasing the number of galaxy pairs within the contributing 2PCF-bins. The number of galaxy pairs inside a 2PCF-bin increases quadratically with $n_\mr{gal}$, therefore it would be interesting to test the ring statistics on a data set like e.g. the COSMOS survey. Similarly, an increased survey volume will significantly enhance the constraints, for the reason that the cosmic variance scales with $1/A$. For example, the CFHTLS data we consider here covers an area of $34.2 \,\mr{deg}^2$ with $n_\mr{gal}=13.3$. Testing the ring statistics on the full CFHTLS sample (172 deg$^2$) would be an interesting project in the future. \\

\begin{appendix}
\section{$T_\pm$-functions}
\label{sec:appendix}
In order to define the $T_\pm$-functions used for the calculation of the $EB$-statistics we remap $\vartheta \in [\vartheta_\mr{min};\vartheta_\mr{max}]$ to the $x \in [-1;1]$ and define
\beq
\label{eq:remap1}
x&=&\frac{2 \, \vartheta-\vartheta_\mr{min}-\vartheta_\mr{max}}{\vartheta_\mr{max}-\vartheta_\mr{min}}\,, \\
\label{eq:remap2}
B&=&\frac{\vartheta_\mr{max} - \vartheta_\mr{min}}{\vartheta_\mr{max} + \vartheta_\mr{min}} \,.
\eeq
We choose our filter function $T_+ (x)$ to be the lowest order polynomial which fulfills the two integral constraints of Eq. (\ref{eq:ebfinitecondition2}) and the normalization $\int_{-1}^1 \mr d x \, T_+ (x) \, T_+ (x)=2$. The function reads
\be
\label{eq:Tpeb}
T_+ (x)=\frac{1}{\sqrt Y} \, \left(3 \,B^2-5 -6\,B\,x+3\,(5-B^2)\,x^2\right) \,,
\ee
where
\be
Y=\frac{8\,(25+5\,B^2+6\,B^4)}{5}\,.
\ee
Given the analytic form of $T_+$ the corresponding $T_-$ is uniquely determined through Eq. (\ref{eq:T-fromT+}).
\end{appendix}



\begin{acknowledgements}
The authors want to thank Yannick Mellier and Martin Kilbinger for useful discussions and advise. TE wants to thank Liping Fu for sharing her CFHTLS data and the Insitut d' Astrophysique de Paris for its hospitality during the analysis of the CFHTLS data. This work was supported by the Deutsche Forschungsgemeinschaft under the projects SCHN 342/6--1 and SCHN 342/9--1. TE is supported by the International Max-Planck Research School of Astronomy and Astrophysics at the University Bonn. 
\end{acknowledgements}

\end{document}